\documentstyle[11pt,psfig,paspconf]{article}

\def\gs{\mathrel{\raise0.35ex\hbox{$\scriptstyle >$}\kern-0.6em 
\lower0.40ex\hbox{{$\scriptstyle \sim$}}}}

\begin{document}


\title{The Formation History of Early-Type Galaxies:
    An Observational Perspective}

\author{Richard G. Bower, Ale Terlevich}
\affil{Department of Physics, University of Durham, South Road, 
	Durham DH1 3LE, UK}

\author{Tadayuki Kodama}
\affil{Institute of Astronomy, University of Cambridge, Madingley Road, 
	Cambridge CB3 0HA, UK}

\author{Nelson Caldwell}
\affil{Smithsonian Institution, Steward Observatory, 949 N Cherry Avenue,
	Tucson Arizona 85719, USA}

\begin{abstract}
This talk investigates the formation of early-type galaxies from a 
deliberately observational view point. 
I begin by reviewing the conclusions that can be reached by comparing
the detailed properties of galaxies in present-day clusters, focusing
on the colour-magnitude relation in particular. The overriding
picture is one of homogeneity, implying a remarkable uniformity
in the formation of these galaxies. This picture contrasts with the
increasing activity seen in clusters as a function of redshift,
creating an apparent paradox been the obvious diversity of star
formation histories in distant cluster galaxies and their uniformity in 
local systems. A resolution is feasible so long as star formation occurs
over an extended epoch. 

In addition to placing limits on variations in star formation history  
the existence of a tight `fundamental relations', such as the 
colour-magnitude relation, can be used to investigate galaxy mergers
and to set limits on the degree to which present-day clusters galaxies are
built by combining systems of stars formed in smaller units. The
final part of this talk turns to early-type galaxies in the field,
and tries to apply the same techniques that have been successful in clusters.
This is an emerging field in which appropriate data-sets are only 
just becoming available; however, comparison of the formation histories of
galaxies in a wide variety of environments is key to distinguishing
between the Classical and Hierarchical models for galaxy formation.
\end{abstract}


\section{Introduction}

In this talk, I will review our knowledge of the formation history of 
early-type galaxies. Rather than approaching the problem from the 
theoretical side, we will take the view point of an `observer'. We will 
look at the data and asking what it directly tells us. Which constraints
follow from which aspects of the observations. The situation is
far from hopeless since we know much about the stellar populations
of galaxies in local clusters. Cluster galaxies are, after all, 
relatively simple systems compared to field spirals in which
much  of the past formation history is masked by the present-day
star formation rate. In addition, clusters of galaxies can be observed
at high redshift, providing us with the means of reconstructing the
evolution of their galaxy populations in a statistical sense.

The evolution of cluster galaxies is dominated by simple scaling
relations, and we can use the existence and narrow scatter of these
relations to set powerful constraints on their formation.
I will argue that we have an emerging, if still incomplete, picture
for the evolution of galaxies in clusters. This makes it tempting
to apply the same arguments to galaxies in the field, and I will
briefly compare the properties of early-type galaxies in the clusters
with their counter-parts in the field.

Of course, the picture that we can hope to build from the data alone
is limited: Carlton Baugh will look at the situation from the opposite, 
but complimentary perspective, creating a model derived from applying
simple rules for galaxy formation to the growth of haloes in a 
hierarchical universe. Full understanding of galaxy evolution will
come from a synthesis of these approaches.


\section{Some Paradigms for the Formation of Early-Type Galaxies}

In order to set the context of this discussion, it will first be
helpful to remind ourselves of the some theories that have been
put forward to explain the range of galaxy morphological types. 
I will deliberately concentrate on two alternative scenarios in order
to deliberately polarise the situation. The truth probably lies
in between.

\begin{figure}
\centerline{\psfig{figure=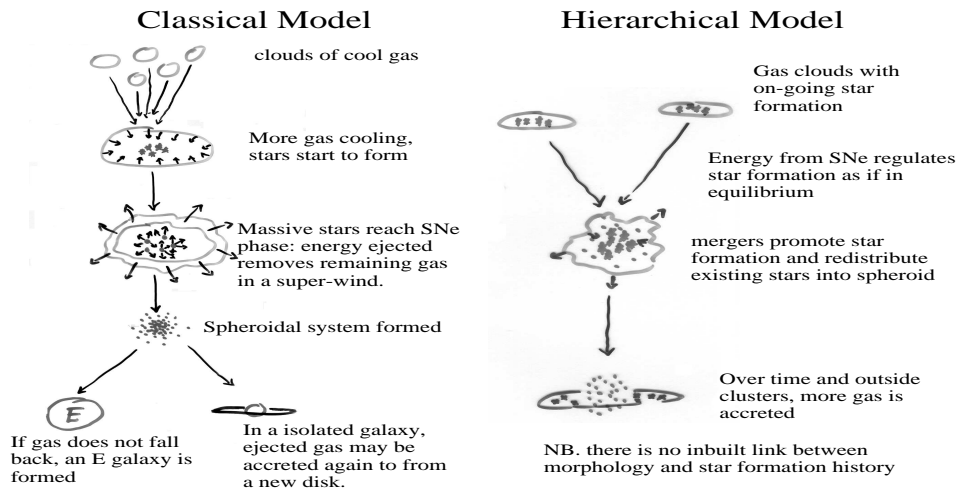,height=2.5in,width=5in,angle=270}}
\caption{A sketch illustrating galaxy formation in the Classical and 
Hierarchical models. In the Classical model, the spheroidal component
is formed in a rapid phase at an early time, and the gas disc is
subsequently accreted over a prolonged period if the galaxy resides
in a low density environment. In the Hierarchical model, galaxies have
no fixed morphology, but can cycle between the different morphological
types depending on the number and strength of interactions and mergers.}
\end{figure}

What I will call the Classical model for galaxy formation has it roots
in the monolithic low collapse models of Larson, 1975. 
However, in my view the monolithic
collapse is not the important component of this scenario: following
the view of Arimoto \& Yoshii (1987), the collapse of a proto
gas cloud may be initially fragmented. Star formation is initially
rapid, and is unchecked until the first massive stars evolve to the 
supernova phase. At this point, star formation can continue only if the
dark matter halo is sufficiently massive to retain the gas against
the increasing pressure of the supernova ejecta. In massive systems, star 
formation will continue longer, but eventually the remaining gas will 
be expelled from the system in a supernova-driven wind, leaving a spheroidal
star system. The subsequent evolution depends on the galaxies environment:
galaxies in clusters loose their wind material to the intra-cluster
medium, while galaxies in low density environments may be able to
reaquire their gas in the form of a gas disc, in which a quiescent mode of
star formation takes place.

By contrast, in the Hierarchical (HGF) model, there is no intrinsic difference 
between star formation occuring at different epochs. 
All galaxies are view as similar
star forming systems in which there is an equilibrium between the
inflow of gas and the rate at which it is either consumed or driven
out of the galaxies by a supernova wind. The morphological appearance
of galaxies is somewhat secondary, resulting from the rearrangement
of stars as the individual sub-components are brought together to make
larger and larger mass units. After each merger, the disk may or
may not grow depending on the whether a gas inflow can re-establish 
itself. This is possible if the galaxy lies at the centre of its new 
halo, but not if it now orbits a larger object.

The key difference between the models is that while morphology is
set at an early time in the Classical model, morphology is a fluid
quantity in the Hierarchical model that can change in both
directions. This division between the Classical and Hierarchical
model parallels the division between the roles of nature and
nurture in forming the morphology of the galaxies.

\section{Galaxies in Local Clusters}

The overriding feature of galaxies in local clusters is their homogeneity:
the uniformity in properties and the way in which they can be scaled 
between galaxies of different luminosities according to a set of `fundamental
relations'.  One key example is the fundamental plane: the correlation
between the luminosity, effective radius and central velocity dispersions
($\sigma$) of early-type galaxies. This seems to have its origin in the 
virial theorem, but also requires considerable uniformity in stellar 
populations (eg., Pahre et al., 1998). Similar strong scalings also
exist between the
Mgb line-strength index and $\sigma$ and between colours and magnitudes
(the colour-magnitude relation: CMR). In this talk I'll concentrate
on the CMR, because this is what I am most experienced in working with,
but the same discussion  could equally apply to these
other relations.

\begin{figure}
\centerline{\psfig{figure=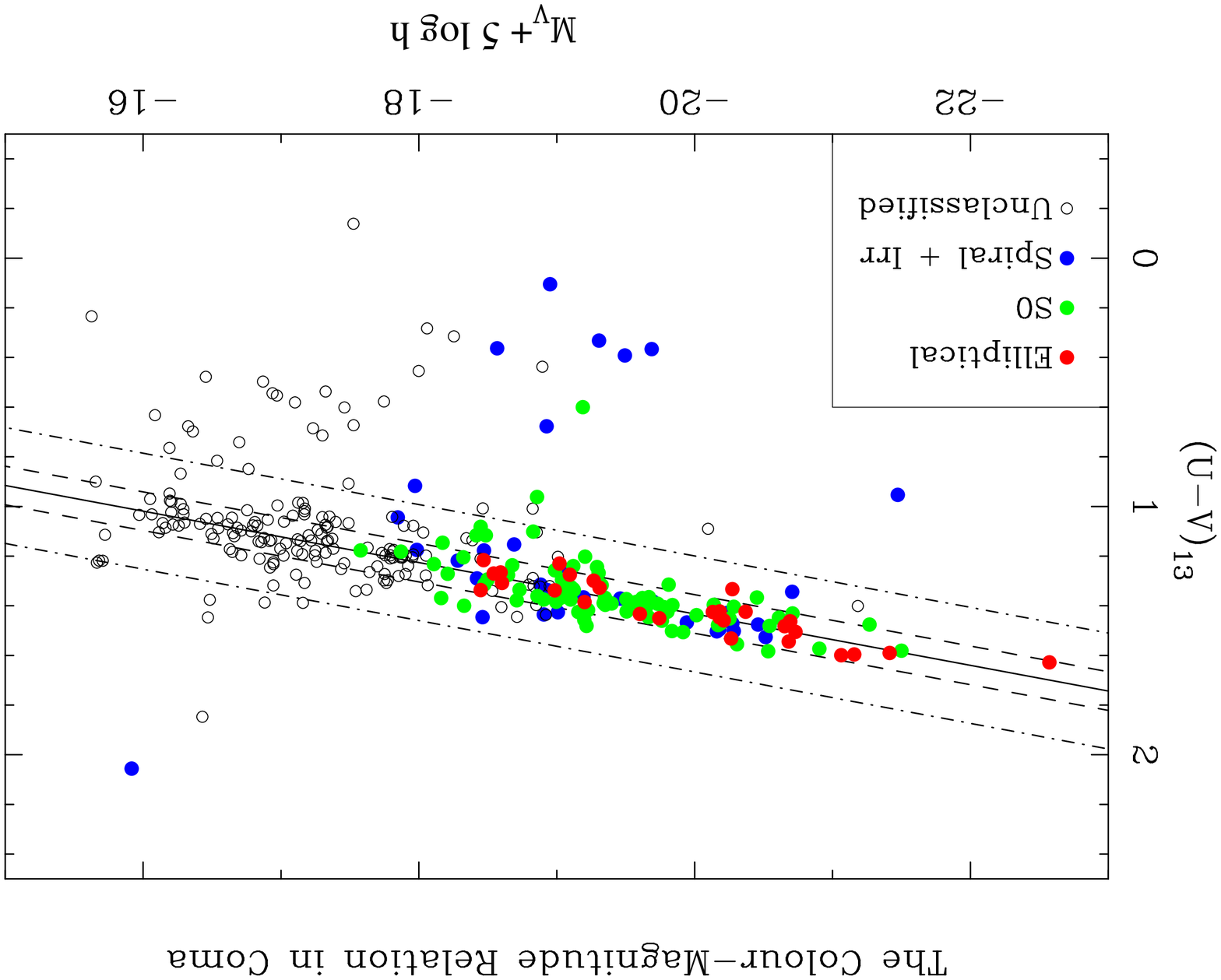,height=2.9in,angle=180}}
\caption{The colour-magnitude relation in the Coma Cluster of galaxies,
from Terlevich et al., 1998. Colours and magnitudes are measured within
an 8.7~kpc aperture. Absolute magnitudes include an average correction
for light outside this aperture. The figure shows the CMR of spectroscopically
confirmed cluster members. 
Symbol colours differentiate E, S0 and S0/a and later galaxies. 
}
\label{fig:comacmr}
\end{figure}

Figure~\ref{fig:comacmr} shows the U-V colour-magnitude relation
for galaxies in the Coma cluster, taken from Terlevich et al., 1998.
The diagram has been limited to galaxies for with spectroscopically 
confirmed cluster membership (mainly from Colless \& Dunn, 1996) and at 
brighter magnitudes, morphologies have been assigned from Andreon et
al., 1996. The over-riding feature
of this diagram is the strength of the colour-magnitude sequence,
extending linearly down to the limit of the spectroscopic catalogues and
beyond.
Although there are a few galaxies lying blueward of the relation, 
most galaxies lie very close the ridgeline. The scatter about the 
ridgeline, measured with a biweight regression estimator (Beers et al., 1990), 
amounts to 0.05~mag for the central $r<20'$ region even if all galaxy
morphological types are included. This can be used to set important 
constraints on the formation history of these galaxies. We will assume
that the overall driving force for the CMR is variations in the
metal abundances of these galaxies, and that the scatter is primarily
due to variations in age. On the basis of the colours alone, this can
only be justified by a plausibility argument; however, the picture
is supported by measurements of line indices in cluster galaxies
(Kuntschner \& Davies, 1998, Terlevich et al., 1998) and by the
evolution of the CMR sequence with redshift (Ellis et al., 1997, 
Kodama et al., 1998, Stanford et al., 1998)

Bower et al., 1992, considered the limits that could be set if the
galaxies were formed in a single burst of star formation. Immediately 
after their creation, galaxies are extremely blue. They become
redder, at first quickly, then more slowly after a few Gyr, until
they asymptotically redden to the colours of the reddest galaxies
in the sequence. The narrow scatter implies either that all the 
galaxies are formed in a well coordinated single event, or that
the galaxies are old enough that the U-V colour is evolving only slowly.
Using stellar population synthesis models to quantify this argument,
Bower et al., found that star formation would need to been completed
before a look-back time of around 10 Gyrs, corresponding to a median
formation redshift of $z\sim2$.

This is a powerful argument, but how is it affected if we assume that
star formation takes place over a more extended period? First it is
necessary to understand why a single burst model is appealing. The key
is to generate a well defined correspondence between the mass
of a galaxy and its metal abundance. In Arimoto \& Yoshii's model
of early-type galaxy formation, this occurs because of the competition
between the tendency of star formation to expel the gas from the 
galaxy in a supernova-driven wind, and the galaxy's gravitational 
potential that keeps the gas trapped. The larger the galaxy,
the longer star formation is able to continue before the gas escapes.
If star formation continues for longer, a large fraction of gas is
consumed and the mean metal abundance of the stars reaches higher values.
This process can only be effective if the star formation period is
short lived.

Hierarchical galaxy formation models require an alternative explanation
for the mass-metallicity relationship. The chemical evolution of a 
closed box of gas tends to a final abundance that depends on the
duration of star formation (Tinsley, 1980). Adding inflow to the systems 
means that an asymptotic metallicity is reached, but this depends
only on the stellar `yield' (ie., the mass of metals returned to the 
interstellar medium for each solar mass of stars that are formed).
The advance made by the HGF models is to consider both the 
inflow and outflow of gas and metals. An essential part of these models
is that the outflow is stronger (and star formation less efficient)
in low mass systems. A side product is that the `effective yield' is
also dependent on mass --- in low mass systems, the metals produced
in supernovae tend to be expelled from the galaxy rather than being
incorporated into the ISM (White \& Frenk, 1991, Kauffmann \& Charlot, 1998).
In this situation a mass-metallicity relation is expected even if the
star formation occurs over an extended period.

\begin{figure}
\centerline{
	\hbox{\psfig{figure=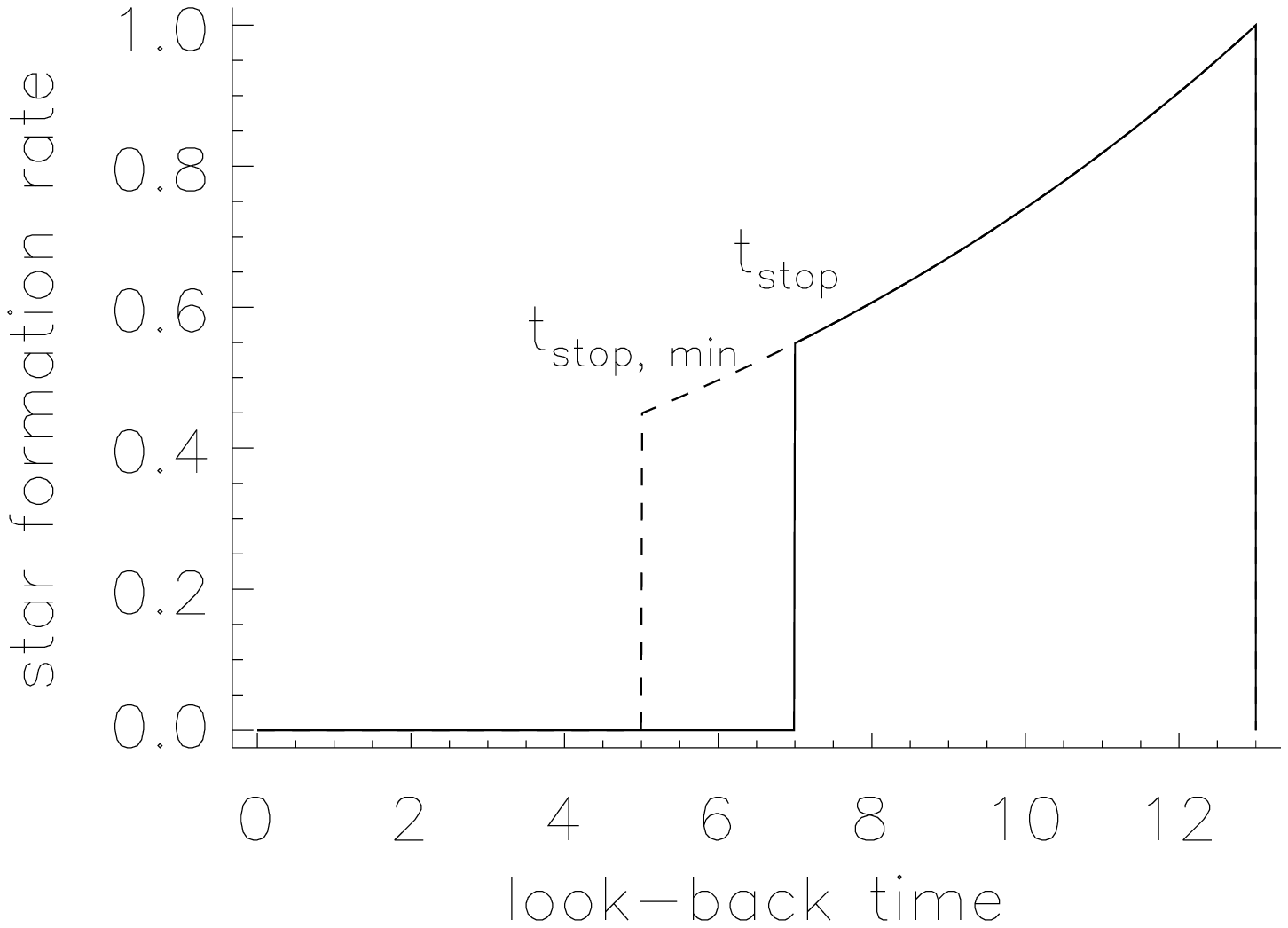,width=2.5in}}
	\hbox{\psfig{figure=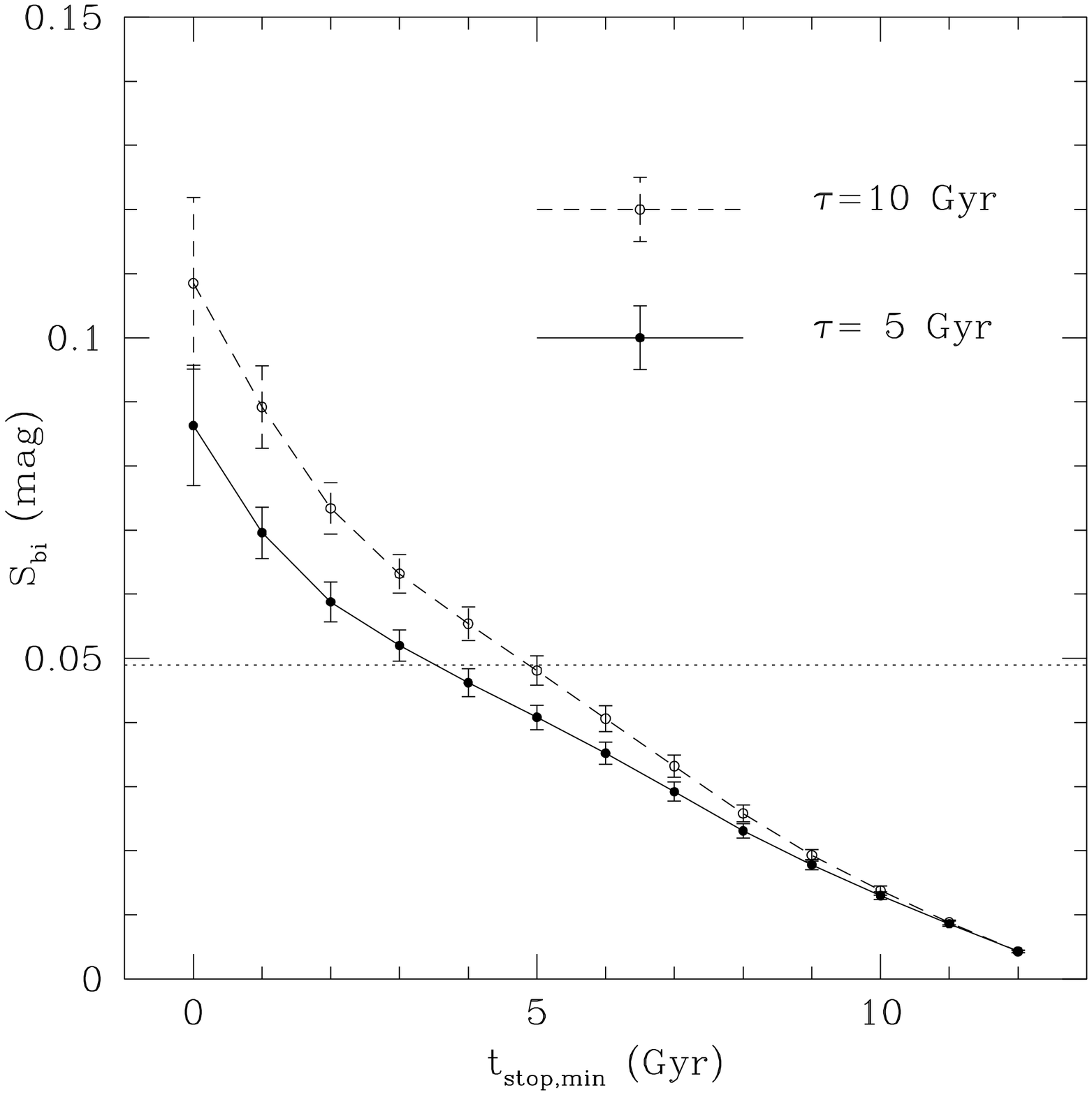,height=2.5in}}
	}
\caption{
This figure shows the limits that can be set on the last epoch 
of star formation from the homogeneity of the local galaxy population.
(a) An illustration of the type of star formation history
considered in the BKT model. (b) The CMR scatter as a function of
the last star formation epoch $t_{stop,min}$. Acceptable models must
lie below the horizontal dashed line.
}
\label{fig:bktscat}
\end{figure}

A single burst model for the formation of early-type galaxies is thus
far too restrictive. In Bower, Kodama \& Terlevich (1998) we have 
explored a wider range of star formation histories. We adopted
a decaying star formation rate ($\tau=5$ or 10 Gyr), and assumed
that all galaxies commence star formation at an early epoch (for our
adopted cosmology this corresponds to a look-back time of
13 Gyr). We modelled the effect of galaxy infall into the
cluster by cutting of star formation at a look-back time $t_{stop}$ that
we assume to be randomly distributed between 13~Gyr and minimum
$t_{stop,min}$ (Figure~\ref{fig:bktscat}) that we adjusted to match the observed scatter in
the Coma cluster CMR relation (using the same bi-weight estimator
for the model and the real data). 

In this model, star star formation is allowed to continue in some
galaxies until recent epochs: for $\tau=5$~Gyr, $t_{stop,min} = 3$~Gyr,
for $\tau=10$~Gyr, $t_{stop,min} = 5$~Gyr. This is considerably less
restrictive than we found for the single burst case, but it is essential
to emphasise that the bulk of the stellar population is still old.
The statement that clusters are dominated by old stellar populations
remains true, but this does not preclude finding relatively young
populations in a small fraction of galaxies.
This is particularly important if we consider the evolution of the 
observed galaxy populations of clusters as a function of redshift.

\section{The Galaxy Populations of Distant Clusters}

Detailed studies of local clusters is one way to reconstruct the 
evolutionary histories of cluster galaxies, but observing clusters
at high redshift, and hence cosmologically significant look-back
times, is a more direct approach. With the Hubble Space Telescope,
observations can be made with comparable resolution and accuracy to
the ground-based observations of the Coma cluster.
Even in the most distant systems studied (eg., $z=1.27$, Stanford et al., 
1997) the CMR still exists. Its evolution is well described by
the anticipated passive evolution of an old stellar population
(eg., Kodama et al., 1998), and even the scatter of morphologically
selected early-type galaxies shows little increase over local
clusters. These observations fit in well with the classical picture
of uniform old stellar populations; however, this passive evolution
does not provide a complete picture of the evolution of galaxies
in clusters.

The key counter-point is the Butcher-Oemler effect (Butcher
\& Oemler, 1984). Their surprising result was that the fraction of
galaxies lying blueward of the colour-magnitude relation increased
dramatically with redshift, from an average fraction of a few
percent in local clusters to $\sim25$\% at $z=0.5$. Although initially,
criticised for the accuracy of the field corrections that needed 
to be supplied, the increase in the "activity" of cluster galaxies
has subsequently been confirmed in spectroscopic studies (eg., Dressler
\& Gunn, 1983, Couch \& Sharples, 1987, van Dokkum et al., 1998,
Poggianti et al., 1998, but see Balogh et al., 1998), and also
in the morphological distribution of cluster members 
(Dressler et al., 1997, Couch et al., 1998). 

\begin{figure}
\centerline{\hbox{\psfig{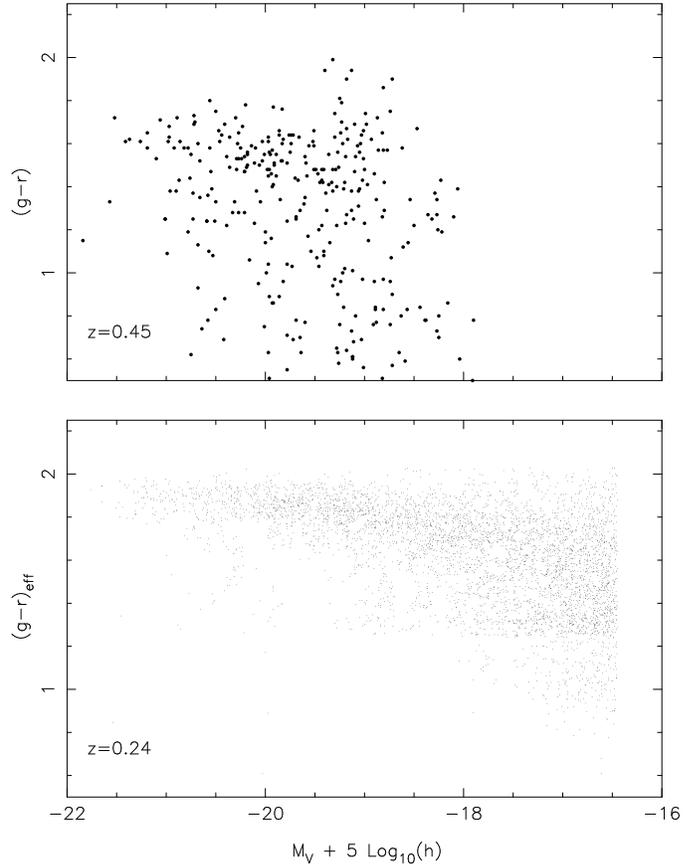}}}
\caption{
A comparison of the CMR in  clusters at $z=0.45$ and $z=0.24$. 
This data should also be compared with the CMR of the Coma cluster
(Figure~\ref{fig:comacmr}). Note both the increase in the number of galaxies
lying blueward of the CMR in the most distant clusters, and the increase
in their characteristic magnitude. This is the Butcher-Oemler effect seen
in the colour-magnitude plane. The data for $z=0.24$ clusters is taken
from Smail et al., 1998.  Data for $z=0.45$ clusters is from the Morphs
collaboration (Dressler et al., 1998). The large scatter about the CMR
is due to the accuracy of the ground-based photometry (cf., Ellis et al.,
1997).
}
\label{fig:hizcmr}
\end{figure}

The strength of the Butcher-Oemler effect is well illustrated
by Figure~\ref{fig:hizcmr}. This compares the CMR in intermediate
redshift clusters (from the Morphs collaboration, Smail et al., 1997,
Dressler et al., 1998) with the CMR for galaxies in rich X-ray selected 
clusters at redshift $z=0.24$ (Smail et al., 1998). 
In both cases, colours and magnitudes have been transformed to the same 
rest frame system. The high redshift data shows only galaxies whose cluster 
membership has been confirmed by spectroscopy, while the lower redshift
data has been field corrected on a statistical basis. The figures should
also be compared to the Coma cluster as a local calibrator. In each
case, the CMR ridgeline is well defined (the larger scatter in the highest
redshift data results from the ground based photometry that has been
used to compile the diagram). 

The striking difference between the diagrams is the increase in both the 
fraction and absolute magnitudes of blue galaxies in the $z\sim 0.4$ dataset. 
This is the Butcher-Oemler effect, but it seems that the effect is
both in the fraction of objects, and in their luminosity. This
maybe related to the phenomenon of `down-sizing' (ie.,
the active population of objects moving to lower intrinsic luminosities
at lower redshifts) that has been observed in field galaxy samples
over this redshift range (eg., Cowie et al., 1997, Guzman et al., 1997). 
It  is important to emphasise that this is not simply a result
of the density--morphology relation (Dressler, 1980), and that the
blue galaxies are found in the cluster cores. It is worth a note
of caution, however, that the blue galaxies may be seen in the
cluster core only due to projection effects (eg., Morris et al., 1998): 
this has still to be factored into the analysis.

\section{Local and Intermediate Redshift Studies in Conflict}

While local clusters present a picture of uniformity and regularity
in star formation history, the distant clusters show evidence of
considerable activity. How can these two sides be compatible?
The galaxies seen in the distant clusters cannot escape them,
and (in a statistical sense) their descendents are found among
the galaxies of present-day clusters that appear so uniform.

The solution to this apparent paradox lies in the freedom to vary
the last epoch of star formation so long as the bulk of the stellar
population remains old. In Bower, Kodama \& Terlevich, 1998 (BKT), we
explored the feasibility of this approach, using the truncated star formation
models discussed in \S2 as our starting point. Taking a random distribution
of truncation epochs (spread over look back times ranging from 0-13 Gyr), 
produces an $\sim50$\% fraction of actively star forming galaxies at 
$z=0.5$. This is more extreme than the canonical Butcher-Oemler effect,
so we considered a variant model in which the cluster was composed
of a 50/50 mix of galaxies with an extended star formation history
and galaxies that are intrinsically old, forming all of their stars
above a look back time of 10~Gyr. This produces a good match to
the observed Butcher-Oemler effect. We included a 10\% burst of star
formation at the moment of truncation in order to reproduce the spectral
features of E+A galaxies.

It is of course questionable whether the random truncation model is
reasonable. In order to produce a more physical motivated model,
we used the extended Press Schechter theory to generated that
infall history of a typical cluster. The mathematical background
to this model is discussed in Bower, 1991. The resulting curve
depends on a the mass scale at which it is assumed that cluster
processes truncate star formation in the accreted galaxies. We
chose the mass scale of a large group for this calculation, although
the results are not sensitive to this assumption. 
Again, if this infall model is used on its own, it tends to over produce the
Butcher-Oemler blue fraction at $z=0.5$, so a variant mixing
truncated and intrinsically old star formation histories was also
considered. 

These models are normalised to match the observed Butcher-Oemler
blue fractions at $z=0.5$. We continued the star formation history
forward in time in order to compare with the CMR cross-section at
low redshift. The scatters we determined from this high idealised model
are shown in Table~\ref{tab:botab}, and should be compared with the scatter of
$\sim$0.05~mag that we measured in the Coma cluster. Neither of
the pure truncation models is able to reproduce a sufficiently narrow
observational relation, although the match to the PS model
is better. However, the models that mix the infalling population with
the intrinsically old component are successful. In these cases, the
reddening of the Butcher-Oemler galaxies is sufficiently rapid that
a well confined present day CMR is recovered. There is not much room
for additional sources of scatter, however.


\begin{table}

\caption{Bi-weight scatter in simulated present-day colour distributions
for a range of star formation histories compatible with observations
of the Butcher-Oemler effect. Model~1 assumes that all cluster galaxies
under go truncation of their star formation at a random time between $t=0$ and 
13~Gyr. Model~2 mixes a 50\% population of these galaxies with a 50\% 
population of galaxies that cease star formation between 10 and 13~Gyr.
Models 3 and~4 are similar except is similar to Model~1, except that the
distribution of truncation times is taken from the infall rate given by 
Bower (1991).}

\begin{center}
\begin{tabular}{lrrrr}
\noalign{\medskip}
\hline\hline
\noalign{\smallskip}
Model&  $\tau=5$~Gyr&  $\tau=10$~Gyr\\
\noalign{\smallskip}
\hline
\noalign{\smallskip} 
 (1)&	0.122 $\pm$ 0.018 & 0.136 $\pm$ 0.021\\
 (2)&	0.052 $\pm$ 0.009 & 0.054 $\pm$ 0.010\\
 (3)&	0.075 $\pm$ 0.010 & 0.085 $\pm$ 0.012\\
 (4)&   0.052 $\pm$ 0.009 & 0.059 $\pm$ 0.010\\
\noalign{\smallskip}
\noalign{\hrule}
\noalign{\smallskip}
\end{tabular}
\end{center}

\label{tab:botab}
\end{table}

Nevertheless, whilst the simple model that we have presented is encouraging 
it is not satisfactory. Two competing process have not been modelled
in adequate detail. Firstly, the galaxies that are counted in the
Butcher-Oemler blue fraction are only those galaxies located in the
core of the cluster (within the radius containing 30\% of the total
galaxy population). Our model assumes that the cluster is static,
and does not allow for galaxies that are in the outer parts of the
cluster at intermediate redshift becoming incorporated into the cluster 
center at the 
present day. One way to deal with this might be to consider galaxies
out to a larger radius in the distant clusters, compared to the
local systems. Allowing for this effect would tend to make
it considerably harder to reconcile the local and intermediate redshift
colour-magnitude diagrams. Carefully modeling through N-body simulations is
required to adequately calibrate this effect, however. For example, 
Morris et al.\ (1998), argue that the Butcher-Oemler galaxies in the cluster
MS1621.5+2640 are located on the periphery of the cluster in velocity
space. 

Secondly, the model is inaccurate because it takes no account of the
fading of galaxies after the truncation of their star formation.
This will tend to remove the fading galaxies from a magnitude
limited sample, helping to make the intermediate and low redshift
data more compatible. For galaxies with simple truncated star formation
the effect is expected to be $\sim 1$ mag if they are selected in
blue light. The mismatch could be minimised by selecting at near-infrared
wavelengths, however.



\section{The CMR as a Constraint on Growth through Mergers}

In addition to setting a limit on the star formation histories
of galaxies, the CMR can also be used to set a limit on the
number of mergers that a galaxy might have undergone after the
formation of its stars. The limit results from the fact that 
mergers between randomly selected galaxies tend to average the
galaxy colours, reducing the slope of any CMR that is initially present.
The scatter also increases since galaxies under go different numbers
of mergers. Thus, using the observed ratio of the CMR scatter to slope,
we can estimate the number of sub-units that that may have been
combined to from a present-day massive elliptical galaxy.
It is important to realise, however, that this argument cannot set
a limit on the number of gas-rich components that may have been 
combined, and subsequently formed new stars.

The basic procedure we use is to assign galaxies to an initially
scatter-free CMR. The actual slope of the relation is not important.
Galaxies are then selected to be merged together. In BKT, we
investigated two models: one in which galaxies were selected at
random, and another in which the galaxies were selected according to
a hierarchical merging tree - as used in HGF simulations (eg., Baugh
et al., 1998). In the random case, just a few mergers are enough
to erase the memory of the initial CMR and to result in a relation
that is incompatible with the observed one. The effects in the
hierarchical merging case are more subtle, however, with galaxies
of equal mass being more likely to merge together at large look-back
times, and unequal mass mergers dominating at later times. 

\begin{figure}
\centerline{
	\hbox{\psfig{figure=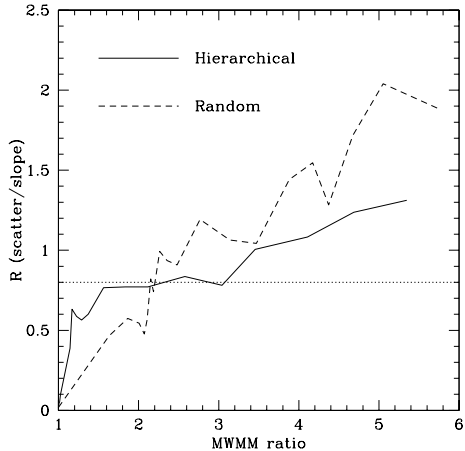,height=2.5in}}
	}
\caption{The growth of scatter in the CMR as a function of the
factor by which the mass of an average galaxy has grown. The parameter
$R$ is the ratio of the scatter to slope in the resulting colour-magnitude
relation. 
}
\label{fig:merger}
\end{figure}

In terms of a time sequence the growth of the scatter:slope ratio ($R$) is 
very different for the two cases. However, the results are 
similar if $R$ is plotted as a function of
the factor by which the mass of a typical galaxy has grown. This
is illustrated in Figure~\ref{fig:merger}, and compared to the
observed ratio (taking into account the aperture in which the CMR
colours have been measured). In both cases, we
obtain similar limits: the observational ratio requires that the
mass of these galaxies cannot have grown by more than a factor
of 2--3 after the formation of the bulk of the stellar population.
This is quite a stringent limit, even though it assumes that the whole of the
scatter is contributed by the merger process, and makes no allowance 
for the scatter resulting from differences in stellar populations.

\section{The Evolution of Field Galaxies}

As I have outlined in the previous sections, the existence of the CMR 
in clusters of galaxies has allowed us to learn a great deal about the
evolution of galaxies in dense environments. It is of course
tempting to apply the same techniques to Elliptical and S0 galaxies in 
lower density
environments: by sampling only galaxies in clusters, we have only
looked at the formation histories of a small fraction of the
early-type galaxy population. 

However, caution is needed. Within cluster cores, most galaxies
have early-type morphology, and, at low-redshift, we obtain very similar 
results regardless of whether a morphological filter is applied or not.
Indeed, we have argued that it is preferable to disregard morphology
completely if we are to use observations of intermediate redshift
clusters to reconstruct the full picture of star formation. In the
field, our results will obviously be sensitive to the criterion
by which the galaxies are selected. Early-type galaxies from only
a small fraction ($<20$\%) of the field galaxy population. Therefore by
selecting galaxies of one particular morphological type, we have
strongly influenced or results and need to be very careful about how we 
are biasing our answers through this selection process.

\begin{figure}
\centerline{
	\hbox{\psfig{figure=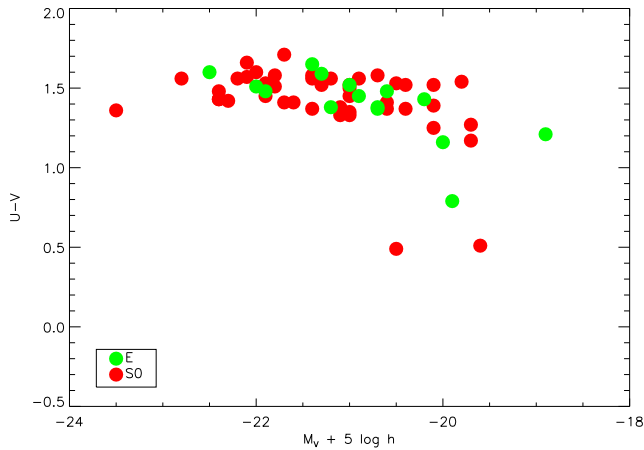,height=2.5in}}
	}
\caption{
The CMR of early-type galaxies in Hickson Compact groups. The figure
has been made by combining the data for many separate groups from
Zepf et al., 1991. The CMR is similarly well defined as in rich clusters.
Although there is a suggestion that the relation maybe shallower
and have large scatter, the differing observational approaches need to
be carefully allowed for. 
}
\label{fig:hickson}
\end{figure}

A first stopping point, could be the CMR in Hickson Compact Groups.
A composite of many different groups is shown in Figure~\ref{fig:hickson}
from Zepf et al., 1991. 
In these systems, we might have expected to find considerably
less uniformity than in the rich clusters because the short
dynamical timescale in these systems makes them a ripe environment
for morphological transformations. Yet, the CMR is remarkably similar
to that seen in the Coma cluster. Possibly it is slightly shallower,
(however the colours are total rather than aperture values) but the 
overall appearance is similar and the scatter is relatively small,
especially allowing for the difficulty in achieving homogeneous photometry
across a wide variety of systems. 
Perhaps the environment selected by the Hickson groups is not
so different to the clusters: there is some evidence to suggest
that we are seeing the cores of much larger mass concentrations. 
The properties of early-type galaxies in compact groups may not, therefore, be
representative of the true `field'.

Data on the CMR of genuine field galaxies is remarkably scant. Most
redshift surveys reach magnitudes at which morphological classification
becomes unreliable, and it is extremely difficult to accurately tie
together the calibration of data-sets covering small numbers of galaxies.
One of the most comprehensive studies remains that of Larson, Tinsley \&
Caldwell (1980). Their data suggested that the CMR for elliptical 
galaxies showed considerably larger scatter in the field, indicating
a wide diversity of star formation histories. Surprisingly, however,
the CMR for S0 galaxies did not show the same variation. And independent
confirmation of their result is clearly needed.

\begin{figure}
\centerline{
	\hbox{\psfig{figure=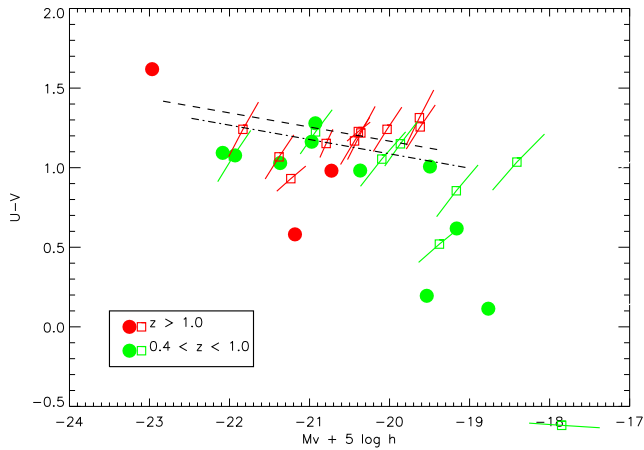,height=2.5in}}
	}
\caption{
The CMR of early-type galaxies in the Hubble Deep Field (Kodama, Bower
\& Bell, in prep).  The galaxies have been corrected both to the
rest frame filter, and have been corrected for the passive evolution
between the true redshift and the median redshift of the sample ($\bar z=0.9$).
The dashed line illustrates the position of the coma cluster CMR at
$z=0.9$ corrected for passive evolution.}
\label{fig:hdfcmr}
\end{figure}

Perversely the best field data may come from substantially higher redshifts. 
The Hubble deep field provides an opportunity to study a sample of tens 
of early-type galaxies with deep, well-calibrated uniform photometry. 
Franceschini et al., 1998,  have used this data-set to study the evolution 
of early-type galaxies. However, the red sequence is also very
well defined. Only a small subset of the early-type galaxies have
confirmed spectroscopic redshifts (Cohen et al.\ 1996), 
however, photometric redshifts
should be reliable for these systems since they poses a well defined
4000\AA\ break (Kodama, Bell \& Bower, 1998). Figure~\ref{fig:hdfcmr}
shows the CMR reconstructed in this way. Many of the blue ellipticals
show morphological peculiarities suggestive of recent interactions.
By contrast, the majority of the galaxies have red colours adhering
quite closely to the CMR defined by the Coma cluster. The scatter in
the relation is quite large, our best estimate indicates 0.11~mag
after allowing for the uncertainties in the K-corrections; however, 
the galaxy sample has a median redshift of $z=0.9$, and the formation
redshift for the stars in many of these systems seems to lie
above $z=2$. Although, the effects of the morphological
filter need to be carefully incorporated, it is likely that this
data will be a significant hurdle for the formation of early-type
galaxies through a continuous transformation of morphology. On the
other hand, the evolution of K-band number counts (eg., Bershady et al.,
1998) and the space density of early-type galaxies at $z>1.5$ 
(Franceschini et al., 1998)
suggest a decline in the numbers of bright early-type galaxies at high
redshifts. It is not clear how these results will be reconcilled, but
dust obscuration and galaxy merging may be crucial factors.

\section{Conclusions} 

What have we learned from this brief tour? This list attempts 
to provide a brief summary.

\begin{itemize}

\item  The stellar populations of galaxies in local clusters are
predominantly old. However, this does not preclude some of the stars
in some of the systems having being formed at relatively recent
epochs.

\item
Distant clusters show passive evolution of the red sequence. But this
is not the complete picture. There is a strongly evolving population
of actively star forming galaxies.

\item
The star forming galaxies in distant clusters must be included in the
progenitors of the red homogeneous population seen in local clusters.
A simple model for infall into the clusters appears to balance these
opposing constraints, but is over simplified. A more complete model
is needed that takes into account the radial distribution of the 
galaxies and their fading from magnitude limited samples.

\item
CMR also places a limit on merging of galaxies. Luminous early-type
galaxies in clusters cannot, on average,  have grown by a factor of more 
than 2-3 in mass after the formation of the bulk of their stars.

\item
We finally considered the CMR for galaxies in the field. Modern
data is surprisingly sparse, and our best results have come from
the Hubble Deep Field using galaxies with a median redshift of
$z=0.9$. Even at this redshift, in low density environments, the 
CMR is well defined and has quite small scatter.
\end{itemize}

At the beginning of this talk I deliberately polarised the discussion
of early-type galaxy formation scenarios into two view points: the
Classical model and the Hierarchical model. Can we now choose between
the two?  Initially, the small scatter of the cluster CMR seems to
favour the Classical model. However, the situation is not quite this
simple since clusters of galaxies --- especially the rich ones that
have been the primary targets to date --- are special places in the universe
in which galaxy formation is advanced with respect to the background
cosmos. For example, Lyman break galaxies at $z\sim 3-4$ are strongly
clustered and seem to be destined to become the core population
of today's clusters (Governato et al., 1998). In these regions the two 
models are not as different as the seem at first sight --- indeed 
Kauffmann \& Charlot (1998) showed that the narrow CMR could be reproduced
in the HGF model. 
The key difference between the models lies in the prevalence of old 
stellar populations, and in the way in which morphology is tied to stellar 
age. The best tests will come
from galaxies in field regions and lower density clusters, and from
galaxies in clusters at high ($z\gs1$) redshifts. In these regions
the differences should become more pronounced as the biasing effect
of the selected structure becomes weakened. Nevertheless, detailed predictions
are required for both models in order that our resources can be
efficiently targeted.

\section*{Acknowledgements}

Thanks are due to Carlton Baugh for provision of the hierarchical
merging trees,
and to Ian Smail and Seve Zepf for their help in producing the colour 
magnitude diagrams of distant clusters and groups respectively.

\begin{question}{Carlton Baugh}
The Classical and Hierarchical models may predict similar properties for
cluster ellipticals today. However, at high redshift the hierarchical 
model would predict these galaxies to be in small fragments, while in the
classical picture the should have similar size. Can this be used to
distinguish the models?
\end{question}
\begin{answer}{Richard Bower}
Yes - that's an important consideration, and there does seem to be some
evidence to support the hierarchical viewpoint from the redshift
distribution of red galaxies and from the evolution of the K-band number
counts (eg., Bershady et al., 1998, Kauffmann \& Charlot, 1998, MNRAS,
297, 23).
These studies suggest that there are too few brighter red galaxies at $z>1.5$.
The interpretation of this result is more complex however: is it due to
the red galaxies becoming fragmented, or due to them becoming bluer because
of star formation, or fainter because of dust reddening. It's a very
powerful way to explore early-type galaxy evolution. 
\end{answer}

\begin{question}{Mike Pahre}
In the work of Rakos \& Schombert (1995, ApJ, 439, 47), they measured
the blue galaxy fractions in clusters out to $z=1$. Have you looked at 
whether the blue fractions at these high redshifts can be reproduced?
\end{question}
\begin{answer}{Richard Bower}
Rakos \& Schombert find that the blue galaxy fractions go on increasing
beyond intermediate redshifts. This is qualitatively consistent with
what you'd expect for infall: at $z=1$, there is very little time to
put the cluster together before the epoch at which it is observed, so
the infall rate has to be very high. We ran into several problems trying
to make an accurate quantitative comparison though, its hard to make
sense of the non-standard filter systems especially. It's also true
even the most distant clusters can have low blue fractions (eg.,Stanford
et al., 1997).  
\end{answer}

\end{document}